# Fast digital refocusing and depth of field extended Fourier ptychography microscopy

GUOCHENG ZHOU,[1,2] SHAOHUI ZHANG,[1,2,*] CHUANJIAN ZHENG,[1] TONG LI,[1] YAO HU,[1] AND QUN HAO[1,*]

[1]*School of Optics and Photonics, Beijing Institute of Technology, Beijing 100081, China*
[2]*These authors contributed equally to this work*
*\*qhao@bit.edu.cn*
*\*zhangshaohui@bit.edu.cn*

**Abstract:** Fourier ptychography microscopy (FPM), sharing its roots with synthetic aperture technique and phase retrieval method, is a recently developed computational microscopic super-resolution technique. By turning on the light-emitting diode (LED) elements sequentially and acquiring the corresponding images that contain different spatial frequencies, FPM can achieve a wide field-of-view (FOV), high-spatial-resolution imaging and phase recovery simultaneously. Conventional FPM assumes that the sample is sufficiently thin and strictly in focus. Nevertheless, even for a relatively thin sample, the non-planar distribution characteristics and the non-ideal position/posture of the sample will cause all or part of FOV to be defocused. In this paper, we proposed a fast digital refocusing and depth-of-field (DOF) extended FPM strategy by taking the advantages of image lateral shift caused by sample defocusing and varied-angle illuminations. The lateral shift amount is proportional to the defocus distance and the tangent of the illumination angle. Instead of searching the optimal defocus distance in optimization strategy, which is time consuming, the defocus distance of each subregion of the sample can be precisely and quickly obtained by calculating the relative lateral shift amounts corresponding to different oblique illuminations. And then, digital refocusing strategy rooting in the Fresnel propagator is integrated into FPM framework to achieve the high-resolution and phase information reconstruction for each part of the sample, which means the DOF the FPM is effectively extended. The feasibility of the proposed method in fast digital refocusing and FOV extending are verified in the actual experiments with the USAF chart and biological samples.

## 1. Introduction

Fourier ptychography microscopy (FPM) [1-9] is a recently developed computational microscopic imaging method, because of its ability to simultaneously achieve large field of view (FOV), high spatial resolution and phase imaging. By sequentially turning on each light-emitting diode (LED) elements located at different positions and stitching the corresponding low-resolution (LR) images together in the Fourier domain, FPM can break through the frequency limit of the employed objective determined by its numerical aperture (NA) and the illumination wavelength. Therefore, the space-bandwidth product (SBP) of the optical imaging system can be effectively increased without any precision mechanical scanning. Nevertheless, there is a critical assumption in conventional FPM framework, which requires the sample under test is sufficiently thin and strictly in focus. In general, the thin sample assumption is considered to be consistent with the first Born approximation condition ($kt\delta n \ll 1$), where $k$ is the wave vector of the illumination sources, $t$ indicates the sample thickness and $\delta n$ is the refractive index difference between sample and medium. Therefore, the theoretical sample thickness in FPM framework should be less than 0.5um for a typical biological sample in air [10]. However, even for a relatively thin sample, the non-planar distribution characteristics and the non-ideal position/posture of the sample will cause all or part of FOV to be out of focus. For example, if a thin sample is three-dimensional curved distributed, at most only part of it can be in focus for any cases. Besides, since precise pose adjustment are difficult to be performed during sample placement, even thin sample distributed on a plane maybe out of focus or in parts (when the sample plane is tilted placed). The above-mentioned defocusing situations will result in evidently quality decreasing of the reconstructed complex high resolution (HR) image. Therefore, the depth of field (DOF) of conventional FPM system is restricted within a very small range, which limits its scope of application.

To deal with these problems, a conventional way is mechanical refocusing, which relies heavily on the accuracy and response time of the mechanical device and will significantly increase the cost of system construction. In contrast, with the help of computational imaging technique, some reconstruction strategies based on FPM have been proposed to realize digital refocusing and extend the DOF of microscopic system. For instance, digital refocusing can be achieved by embedding an optimization search module in the standard FPM algorithm framework. By defining a convergence index and iteratively searching the maximum value corresponding to different defocus distances in FPM scheme, Zi. Bian et al. have realized defocus distance correction and imaging DOF extending for thin sample [11]. Nevertheless, there are a lot of repetitive calculations in the above optimal algorithm, so digital refocusing is a very time-consuming process, especially when the defocus distances for each part are different. Besides, by implementing different 3D sample model in FPM framework, such as multi-slice [12] and 3D k-space model [13], tomography has been accomplished for thick sample, only part of which is in strictly focus. These FPM tomography strategies can also be utilized for thin sample with non-ideal distribution and placement. The FPM based on multi-slice model is also time consuming because a lot of forward and backward propagation steps are involved in its reconstruction algorithm. And, the tomographic algorithms based on 3D k-space model are relatively limited in reconstruction depth, usually on the order of tens of microns.

To increase the digital refocusing efficiency, some new methods have been proposed in terms of system working mode and reconstruction algorithms, respectively. Symmetrical illumination can alleviate the lateral shift of the out-of-focus image to a certain extent, and therefore expand the DOV [14]. Nevertheless, the free propagation of the out-of-focus part has not been reasonably considered in this approach. Furthermore, symmetrical illumination can also cause further blurring of the image, when partially eliminating the effect of image shift. In terms of algorithms optimization, different from Bian et al.'s scheme of embedding optimal module within standard FPM scheme, R. Claveau et al. realize DOF extending and digital refocusing successfully by propagating the reconstructed HR complex images to different defocus distances and finding the optimal focus position according to an image processing pipeline [15]. However, the image processing pipeline is still an optimization procedure and requires large amount of calculation. Moreover, directly performing FPM reconstruction on the out-of-focus sample information will have some system parameter deviation problems.

In this paper, we proposed a fast digital refocusing and depth of field extended Fourier ptychography microscopy by taking advantage of image lateral shift caused by sample defocusing and angle-varied oblique illuminations. The lateral shift amount is proportional to the defocus distance and the tangent of the illumination angle. Instead of using optimization strategy, which is time consuming, the defocus distance of each subregion of the sample can be precisely and quickly obtained by calculating the relative lateral shift amounts corresponding different oblique illuminations. Thus, defocus distances corresponding to different subregions can be obtained by calculating relative image shift amounts corresponding different oblique illuminations. And then, digital refocusing strategy rooting in the Fresnel propagator is integrated into FPM framework to achieve the high-resolution and phase information reconstruction for each subregion of the sample, which means the DOF of the FPM is effectively extended. The feasibility of the proposed method in fast digital refocusing and DOF extended is verified in the actual experiments with the USAF chart and biological samples. This paper is organized as follows. The principle of standard FPM framework is presented in Section 2.1. The characteristics of lateral shift for out-of-focus sample under oblique illumination is discussed in Section 2.2. And, the workflow of our method is presented in Section 2.3. In Section 3, the USAF chart and biological samples are used to demonstrate the effectiveness of our method in digital refocusing and DOF expansion. Conclusions are summarized in Section 4.

## 2. Principle

### 2.1 FPM principle and system setup

A conventional FPM system setup is shown in Fig. 1 (a), in which a LED array is utilized for varied-angle illumination. FPM assumes that the sample is sufficiently thin and illuminated with a monochromatic plane wave. Illuminations from different angles will result in different spectrum shift in the Fourier domain. Unlike real-space ptychography that acquires diffraction patterns in the Fourier domain [16, 17], FPM records LR images in the spatial domain directly, which can reduce the requirement of dynamic range of the camera efficiently. By stitching LR images in the Fourier domain, FPM can achieve a large FOV, high-resolution and phase recovery imaging simultaneously. LR images acquisition process can be described as Eq. (1).

$$I_n(x,y) = \left| \mathfrak{F}^{-1}(\mathfrak{F}(t(x,y)) \cdot P(u,v)) \right|^2, \tag{1}$$

Where $t(x,y) = s(x,y) \cdot e^{i(xk_{xn}+yk_{yn})}$ denotes the exit wave distribution of the sample $s(x,y)$ that is illuminated by an oblique illumination with a wavevector $(k_{xn}, k_{yn})$. '$\mathfrak{F}$' and '$\mathfrak{F}^{-1}$' indicate the Fourier and inverse Fourier transform respectively. $P(u,v)$ is the pupil function of the objective, where $(x,y)$ are the 2D spatial coordinates in the sample and $(u,v)$ are the corresponding spatial frequencies in the Fourier domain. $I_n(x,y)$ is the intensity acquired by the camera.

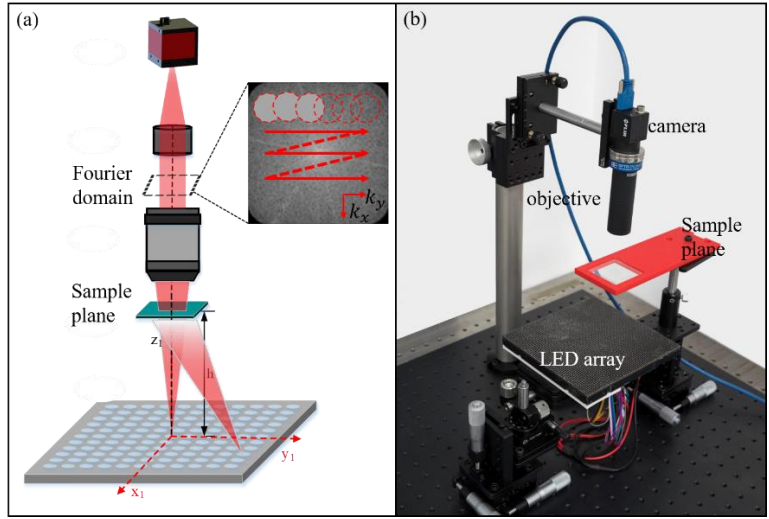

Fig. 1 System setup of FPM. (a) principle of FPM. (b) system setup in actual experiments.

As shown in Fig. 1 (b), we build an FPM system to acquire LR images. We utilize an objective (OPTO, TC23004, magnification 2x, NA≈0.1, FOV 4.25mm×3.55mm, working distance 56mm) instead of conventional microscopes for optical imaging, which simplifies the system construction and owns a longer working distance. A LED array (CMN, P 2.5, 19×19) is used for varied-angles illuminations. The wavelength of illuminations is 470nm. A camera (FLIR, BFS-U3-200S6M-C, sensor size 1", dynamic range 71.89dB, pixel size 2.4μm) is used for recording LR images. In our experiments, the distance between the LED array and the sample plane is set to 83mm. Using the exposure time of 30mm, we acquire 225 LR images for a set of raw data.

## 2.2 Model of image shift

If a sample is placed at out-of-focus plane, there will be an image lateral shift between BF LR images corresponding to varied-angle illuminations. Zheng et al. proposed a single-frame autofocusing hardware for whole-slide imaging system by searching the non-shift position [18]. We found that this image lateral shift is proportional to the defocus distance and the tangent of illumination angle. Different defocus distances result in different image lateral shifts between BF images with varied-angle illuminations [19]. As shown in Fig. 2, a sample placed at out-of-focus plane is illuminated by two monochromatic plane waves with different illumination angles. Where light source $a$ locates at the optical axis, and the angle between light sources $a$ and $b$ is $\theta$. Eq. (2) is a phase shift function where $z$ is the defocus distance between focus and defocus plane. According to Eq. (2), there is non-spectrum shift in the Fourier domain and no pixel shift in the spatial domain between focus and out-of-focus plane if the sample is illuminated by a vertical illumination. The image corresponding to the center LED that locates at the optical axis can be used as a reference image in the image shift calculation. And the relation between defocus distance and image shift can be described as Eq. (3). Where $\delta s_i$ is the image shift between the center and ith BF images. $A_i = \eta_i \cdot \tan\theta_i$ is a constant value decided by the incident angle between the center and ith images.

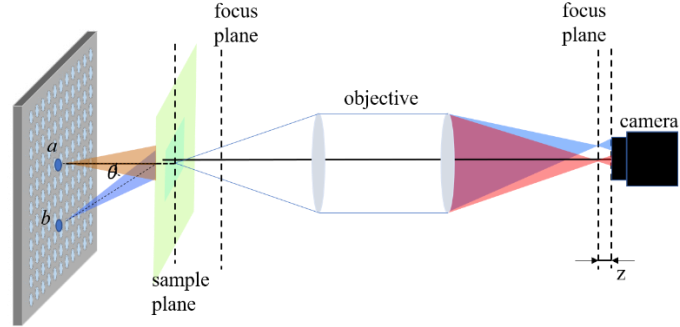

Fig. 2 Imaging model for the lateral shift of the image of a defocused sample under oblique illumination

$$\varphi(k_x, k_y) = z \cdot \tan\theta \cdot (k_x \cos\alpha + k_y \sin\alpha), \qquad (2)$$

$$\delta s_i = \eta_i \cdot \tan\theta_i \cdot z = A_i \cdot z, \qquad (3)$$

According to Eq. (3), we transform calculating the defocus distance $z$ to the lateral image shift $\delta s_i$. To obtain the accurate image shift between images in the spatial domain, several image registration methods can be utilized in this procedure, including Mean Absolute Differences (MAD) method [20], Sum of Absolute Difference (SAD) method [21], phase correlation method [22] and so on. In our method, inspired by Li. Bian et al. [23], we utilize an image registration method which can be expressed as,

$$(\Delta x_i, \Delta y_i) = (x_i - \frac{m}{2} - 1, y_i - \frac{n}{2} - 1),$$
$$\text{and } (x_i, y_i) = \underset{(x_i, y_i)}{\arg\max}\{abs\{\text{fftshift}\{\text{ifft2}[\text{fft2}(I_c(x,y)) \cdot \text{ifft2}(I_i(x,y))]\}\}\}, \qquad (4)$$

where $(\Delta x_i, \Delta y_i)$ are the image shifts corresponding to x-axis and y-axis between the center and ith images respectively. $(x_i, y_i)$ are the coordinates corresponding to the maximum value in the optimal matrix. $m \times n$ is the image size of a LR image. $I_c(x, y)$ and $I_i(x, y)$ are the intensity of the center and ith LR images respectively. Furthermore, although the EPRY algorithm [24] can correct most aberrations of the imaging system, including the offset between LR images, we still should correct this offset before employing LR images in FPM strategy to achieve a correct pupil function of chromatic aberration. Due to the dark-field (DF) images corresponding to high-order diffraction do not exist offsets, thus, only the offsets between bright-field (BF) images should be corrected. In our method, the center image is used as reference, and we calculate the different offsets between the center and other BF images. Then, after obtaining a set of $(\Delta x_i, \Delta y_i)$, we correct the offset of every BF image respectively. Finally, according to Eq. (3), defocus distance can be calculated by the offset. Considering the system misalignments, especially LED position misalignments, we calculate a set of defocus distances $z_i$, and use its average for the final defocus distance in our method.

### 2.3 Digital refocusing scheme

Digital refocusing is an effective method for reconstructing images with different defocus distances, which can be utilized in two ways. Firstly, defocus distance can be classified into an unknown aberration in FPM. By utilizing the EPRY algorithm, defocus aberration can be corrected. However, we found that the EPRY algorithm can only recover defocus aberration in a small range of defocus distances (typically value: -50~50um). As shown in Fig. 3, we perform a simulation to confirm the ability of reconstruction under different defocus distances with the EPRY algorithm and our method, where 225 LR images are utilized and iterations are set to 300. By introducing defocus aberration from 0-300um, we calculate the Structural Similarity (SSIM) between ground truth and reconstruction images with the EPRY algorithm as shown in the blue-lined curve. We found that when the defocus distance is larger than 60um, the quality of EPRY reconstruction result will decrease rapidly. Furthermore, an optical imaging system exists lots of aberrations commonly, including the chromatic aberration, defocus aberration, intensity fluctuation of LED array and so on. The pupil function recovered by the EPRY algorithm is a coupling aberration containing all these aberrations. Thus, the aberration correction process is difficult to obtain the correct objective chromatic aberration, in which the results are also shown in Fig. 5.

Secondly, some digital refocusing methods are used in the final reconstruction result of FPM. By propagating the reconstruction result with a set of certain defocus distances, different images corresponding to different defocus plane can be obtained. Furthermore, REMY et al. also proposed a different free space propagator to lead a higher contrast in refocused images [15]. Nevertheless, as mentioned in [15], this strategy aims to searching the focused plane among a stack of defocus planes by image processing. The EPRY algorithm cannot used in this strategy, because it will auto refocus the reconstruction result to the clearest plane in a small range. And this strategy can be only utilized inside the DOF of objective. When the defocus distance is outside the DOF of objective, the quality of refocused results will decrease rapidly. In our method, the SSIM keeps a higher value even the defocus distance is larger than 250um, which can be found in Fig. 3.

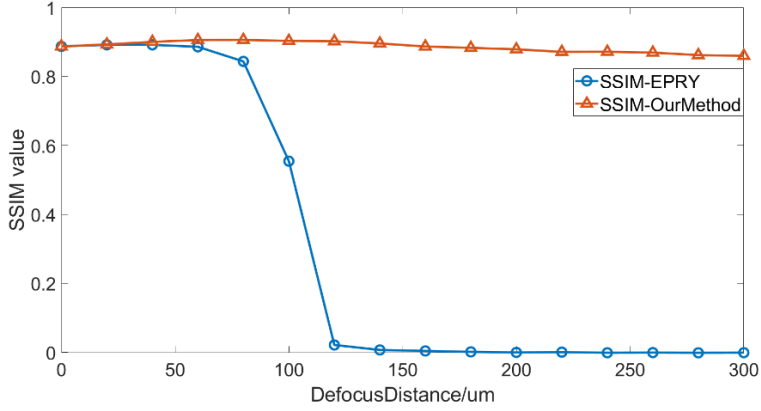

Fig. 3 SSIM value of the EPRY algorithm and our method. The red-lined curve corresponds to our method. The blue-lined curve corresponds to the EPRY algorithm.

In our method, we utilize the Fresnel propagator [25] to realize digital refocusing. The complex distribution of refocused image can be described in Eq. (5).

$$s_1(x,y) = \mathfrak{I}^{-1}\{\mathfrak{I}(s_0(x,y)) \cdot H(k_x, k_y, z)\}, \quad (5)$$

Where $s_0(x,y)$ is a known complex optical field in a given plane $z_0$, and $s_1(x,y)$ is the field in plane $z_1$ obtained by the Fresnel propagation. $H(k_x, k_y, z)$ is a Fresnel propagator described as follows. Where z is the defocus distance between plane $z_0$ and $z_1$.

$$H(k_x, k_y, z) = \exp(j\frac{2\pi}{\lambda} \cdot z \cdot \sqrt{1 - k_x^2 - k_y^2}). \quad (6)$$

In our method, we utilize the digital refocusing method in the FPM iteration scheme. The framework of digital refocusing in our method is shown in Fig. 4.

---

**Algorithm outline**

---

**Input:** a set of LR images; a certain defocus distance z calculating by Eq. (3);
**Output:** recovered object complex distribution $s(x,y) = Ae^{j\varphi}$;

---

Initialize $s_1(x,y) = Ae^{j\varphi}$ at the focus plane, where $A$ and $\varphi$ are the amplitude and phase of object respectively;
$S_1(u,v) = \mathfrak{I}\{s_1(x,y)\}$; $H(u,v,z) = e^{j \cdot \frac{2\pi}{\lambda} \cdot z \cdot \sqrt{1-u^2-v^2}}$;
**for** itera=1:loopNum
  **for** i=1:361
    $O_i(u,v) = S_i(u-u_i, v-v_i) \cdot P(u,v)$;
    $o_{iref}(x,y) = \mathfrak{I}^{-1}[O_i(u,v) \cdot H(u,v,z)] = A_i e^{j\varphi}$;
    replacing the amplitude of $o_{iref}(x,y)$ with the intensity of ith LR image while keep the phase $\varphi$ unchanged;
    $o_{iref\_updated}(x,y) = \sqrt{I_i} e^{j\varphi}$;
    $O_{i\_updated}(u,v) = \mathfrak{I}\left(o_{iref\_updated}(x,y)\right) \cdot H(u,v,-z)$;
    $S_{i+1}(u,v) = S_i(u,v) + \alpha \frac{P_i^*(u,v)}{|P_i(u,v)|_{max}^2} [O_{i\_updated}(u,v) - O_i(u,v)]$;
    $P_{i+1}(u,v) = P_i(u,v) + \beta \frac{S_{i+1}^*(u-u_i,v-v_i)}{|S_{i+1}(u-u_i,v-v_i)|_{max}^2} [O_{i\_updated}(u,v) - O_i(u,v)]$;
  **end**
  $s(x,y) = \mathfrak{I}^{-1}(S(u,v)) = A_{recovered} e^{j\varphi_{recovered}}$;
**end**

---

Fig. 4 Algorithm outline of digital refocusing in our method.

At the beginning of algorithm outline, we initialize the HR reconstruction distribution $s_1(x,y)$ in the spatial domain at the focus plane. The amplitude $A$ can be an all one matrix or initialized by the resized center LR image. The corresponding distribution in the Fourier domain is described as $S_1(u,v)$. We obtain the subregion spectrum $O_i(u,v)$ corresponding to ith LR image according to conventional FPM strategy. By Eq. (5), we propagate ith subregion spectrum corresponding to focus plane onto plane z, and then, obtain a complex distribution $o_{iref}(x,y) = A_i e^{j\varphi}$ at the plane z. Then, the amplitude $A_i$ will be replaced with the intensity of ith LR image acquired by experiments while keeping the phase $\varphi$ unchanged. Finally, propagating the updated complex distribution onto the focus plane and updating the object and pupil functions according to the EPRY algorithm.

Furthermore, we perform a simulation as shown in Fig. 5 to demonstrate the feasibility of our method in recovering the chromatic aberration without defocus aberration. In this simulation, 225 LR images are used in FPM reconstruction, and the total iteration is set to 300. Defocus aberration is set to 200um. Pictures of a 'cameraman' and 'westconcordorthophoto' are used as the ground truth of amplitude and phase of the object as shown in Fig. 5 (a1) and (a2). By using Ornstein-Zernike equations [26], we generate a surface as the chromatic aberration as shown in Fig. 5 (a3). 37 coefficients of surface can be found in Table. 1. By adding this chromatic aberration in LR images, the EPRY algorithm and our method are used to recover the object and pupil functions respectively. Fig. 5 (b1-b3) are the amplitude, phase and pupil function recovered by the EPRY algorithm. Fig. 5 (c1-c3) are the amplitude, phase and pupil function recovered by our method, respectively. Comparing the results shown in Fig. 5 (b) and (c), we found that chromatic aberration and defocus aberration can be recovered effectively by utilizing digital refocusing in FPM optimal processing. While the EPRY algorithm cannot recover the result if the defocus distance is too large.

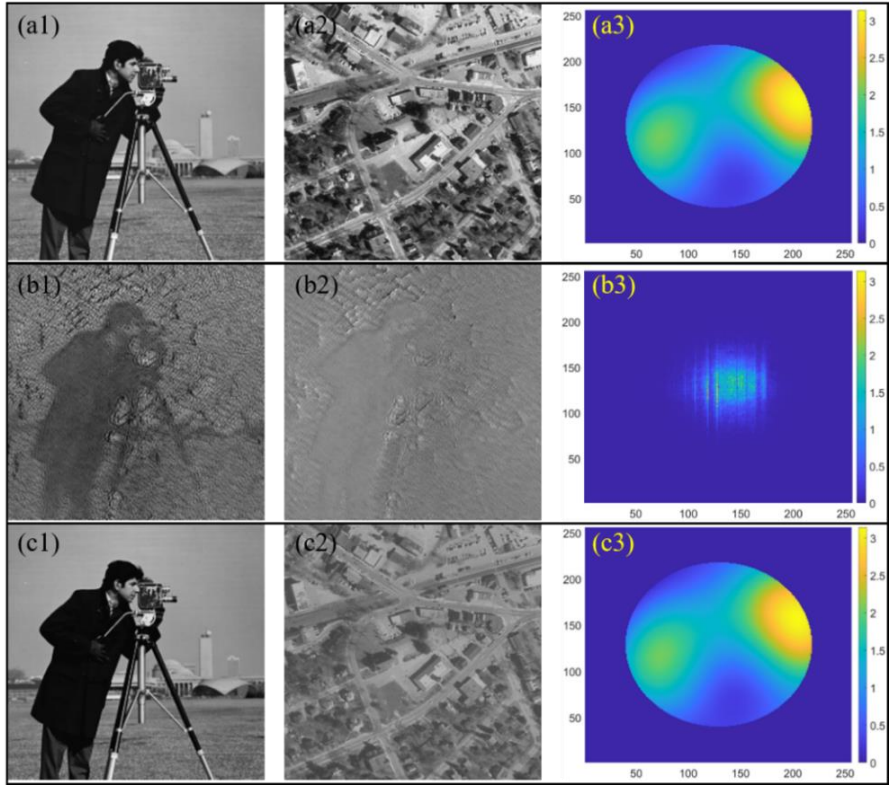

Fig. 5 Chromatic aberration recovery result in FPM. (a1-a2) Ground truth of amplitude and phase of object respectively. (a3) Chromatic aberration generated by Zernike in simulations. (b1-b2) Amplitude and phase of object recovered by the EPRY algorithm. (b3) Coupling pupil function

recovered by the EPRY algorithm. (c1-c2) Amplitude and phase of object recovered by our method. (c3) Pupil function recovered by our method.

Table. 1 Zernike coefficients of chromatic aberration surface

| Coefficients | Z1 | Z2 | Z3 | Z4 | Z5 | Z6 | Z7 |
|---|---|---|---|---|---|---|---|
| Value | 1.3623 | 0.4519 | 1.6484 | -2.0284 | -0.4493 | 0.2360 | -0.8352 |
| Coefficients | Z8 | Z9 | Z10 | Z11 | Z12 | Z13 | Z14 |
| Value | -1.2760 | 0.6170 | 0.6127 | 0.2894 | 0.3953 | -0.8706 | -0.4977 |
| Coefficients | Z15 | Z16 | Z17 | Z18 | Z19 | Z20 | Z21 |
| Value | -0.1067 | -0.6878 | 0.3319 | 2.3652 | -0.4822 | 0.6474 | -1.0344 |
| Coefficients | Z22 | Z23 | Z24 | Z25 | Z26 | Z27 | Z28 |
| Value | 1.3396 | -0.9691 | 0.2087 | -0.6186 | 0.5120 | 0.0114 | -0.0440 |
| Coefficients | Z29 | Z30 | Z31 | Z32 | Z33 | Z34 | Z35 |
| Value | 2.9491 | -0.6300 | -0.0469 | 2.6830 | -1.1467 | 0.5530 | -1.0765 |
| Coefficients | Z36 | Z37 | | | | | |
| Value | 1.0306 | 0.3275 | | | | | |

## 3. Experiments

### 3.1 System scale factor calibration

According to Eq. (3), the coefficient $A_i$ is a constant value that indicates the proportional relation between defocus distance and image shifts of the center and other BF images. Before implementing Eq. (3) to calculate defocus distance, a system scale factor calibration of $A_i$ should be performed firstly. As shown in Fig. 6 (a), we choose the 2rd, 4th, 6th and 8th BF images corresponding to different LEDs to calibrate the scale factor with the center LR image. In our experiment, we use the pure amplitude object USAF target in calibration. By adjusting the sample plane and changing the defocus distance from 0 to -200um, in which step size is set to 20um, we record 25 pictures BF LR images with every defocus distance. Sequentially, we calculate the image shifts between center image and 2rd/4th/6th/8th BF image respectively according to Eq. (4).

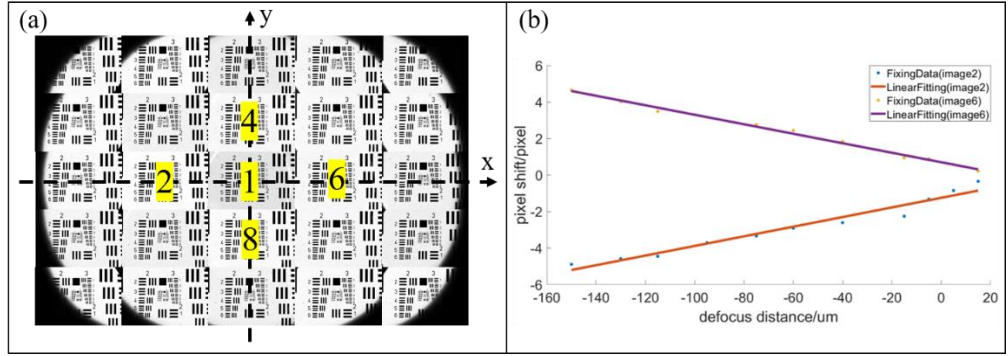

Fig. 6 System scale factor calibration. (a) Space distributions of BR LR images corresponding to different LEDs. The yellow-boxed images are used in calibrations. (b) Linear fitting results corresponding to BF images 2rd and 6th with the center image 1st, where $A_2 = 0.0264$ and $A_3 = -0.0260$.

In our actual experiments, to eliminate the effect from the movement error of moving device, the defocus distance corresponding to each group of raw data is determined by the iteration optimal process shown in [11]. As shown in Fig. 6 (b), we found that the accuracy defocus distances determined by optimal process exist an error with moving device. Then, by using linear fitting method, we obtain the scale factors corresponding to different LR images as follows: $A_2 = 0.0264$, $A_4 = -0.0248$, $A_6 = -0.0260$ and $A_8 = 0.0250$. We also show a set of defocus distance and image shift corresponding to center and 2th image in Table. 2.

**Table. 2 relationship of defocus distance and pixel shift**

| z/um | 15 | 5 | -5 | -15 | -40 | -60 |
|---|---|---|---|---|---|---|
| pixel shift/pixel | -0.35 | -0.85 | -1.33 | -2.26 | -2.61 | -2.91 |
| z/um | -75 | -95 | -115 | -130 | -150 | |
| pixel shift/pixel | -3.35 | -3.71 | -4.46 | -4.59 | -4.90 | |

*3.2 high-resolution reconstruction of USAF target*

High-resolution imaging with a large FOV is a very important characteristic in FPM. To demonstrate the feasibility of HR imaging in our method, the pure amplitude reconstruction is utilized in experiments firstly. As shown in Fig. 7, we use the USAF target to realize HR imaging. In this experiment, the USAF is placed at the sample plane. By changing different defocus distances, we acquire a set of LR images (225 pictures) corresponding to each defocus distance. Then, according to Eqs. (3) and (4), we choose a subregion of LR image as shown in the blue boxed area in Fig. 7 to calculate the image shift between images and defocus distance. To eliminate the positional misalignment of LED array, we use the average of defocus distances corresponding to two couples of symmetrical images in our method. For instant, as shown in Fig. 6 (a), 2rd, 6th and 4th, 8th images are two couples of symmetrical images. The defocus distance can be described as $z = (z_2 + z_4 + z_6 + z_8)/4$, where the subscript number indicates the defocus distance calculated by ith image according to Eq. (3). As shown in Fig. 7, the defocus distances we recorded corresponding to different defocus planes are -192, -101, 0, 95 and 194um respectively.

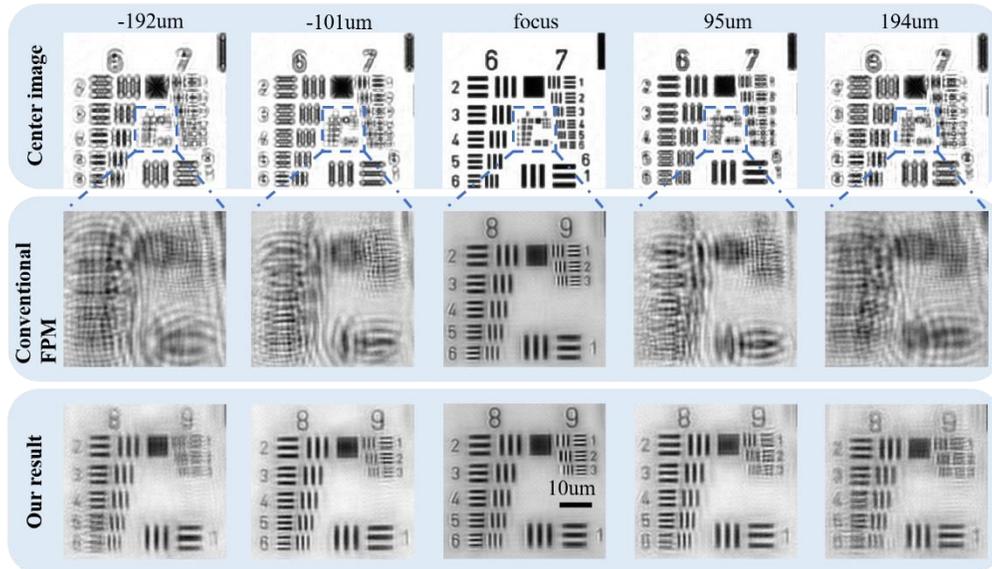

Fig. 7 High resolution reconstruction result with different defocus distances. Conventional FPM is utilized with the EPRY algorithm, while our result is utilized with the proposed method.

Sequentially, by implementing the defocus distance in FPM with our digital refocusing method, the amplitude results of USAF target are shown in Fig. 7. Furthermore, we also use the EPRY algorithm with conventional FPM as comparisons. We found that when the sample is placed at focus plane, both the EPRY algorithm and our method can recover the HR amplitude. With the defocus distance increases, the EPRY algorithm without digital refocusing cannot recover the amplitude effectively, which is consistent with the simulation result shown in Fig.3. In contrast, with our method, the quality of reconstruction can be evidently improved compared with conventional the EPRY method. This results also demonstrate the accuracy of calculating defocus distance with our method. Certainly, in our method, the quality of high frequencies corresponding to lines 8 and 9 with defocus distances -192 and 194um decrease to some extent compared with the quality with focus plane, but still improved a lot with conventional method. We analyze that the high frequencies losing in Eq. (6) will result in the quality difference between defocus result and focus result by using our method.

Furthermore, we also tilt the USAF target and acquire a set of LR images. In this experiment, a subregion of the USAF target is placed at focus plane while other subregions are placed at out-of-focus plane due to the tilted sample, as shown in Fig. 8. Then, three subregions are used to calculate the defocus distance according to Eq. (3) respectively. The defocus distances corresponding to Fig. 8 (a), (b), (c) are 43, 0, -20um respectively. Then, the EPRY algorithm and our method are utilized to reconstruct the amplitude of USAF. As shown in Fig. 8, the x-axis distance between subregions Fig. 8 (a) and (b) is 2250um approximately. Similarly, the x-axis distance between subregions Fig. 8 (b) and (c) is 1125um approximately. Thus, the titled angle of sample we calculated is 1.1approximately.

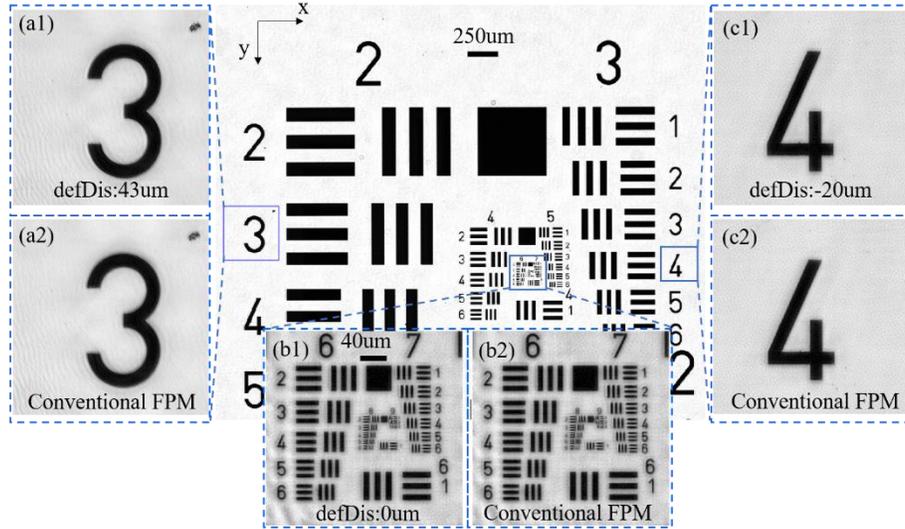

Fig. 8 Tilted USAF target reconstruction result with the EPRY algorithm and our method (a1-a2) Reconstructed amplitude of The EPRY algorithm and our method with 43um defocus distance. (b1-b2) Reconstructed amplitude of The EPRY algorithm and our method with 0um defocus distance. (c1-c2) Reconstructed amplitude of The EPRY algorithm and our method with 20um defocus distance.

### 3.3 Phase recovery of biological sample

Phase imaging is also an important characteristic in FPM. Generally, a biological sample is a 3D plane, different subregions have different defocus distances. In conventional FPM, we assume that the biological sample is placed at the focus plane, which certainly miss lots of important information in different subregions. To demonstrate the feasibility of phase imaging in our method, a biological sample Paramecium is utilized in this experiment. As shown in Fig.

9, we place the sample randomly at the sample plane without adjusting the defocus distance accurately and acquire a set of LR images (225 pictures). The whole FOV image is segmented into several subregions and different defocus distances corresponding to different subregions are calculated according to Eq. (3) respectively. Then, the conventional FPM scheme and our method are utilized to reconstruct the objective function respectively. As shown in Fig. 9, three subregions are used for phase reconstruction. According to Eq. (3), the defocus distances corresponding to three subregions are 79, 80 and 60um respectively. By utilizing conventional FPM and our method, the amplitude and phase are shown in Fig. 9 respectively.

Compare our results with conventional results in subregions 1 and 3, we found that by using digital refocusing method, our method has a more clearer imaging quality. For instance, as shown in the yellow arrows in Fig. 9 (c1), due to defocus imaging with 60um defocus distance, the vignetting appears in the result of conventional FPM. Nevertheless, by using the digital refocusing of our method, the vignetting phenomenon can be improved as shown in Fig. 9 (c3). Furthermore, as shown in the red arrows in subregions 2, several cells are overlapped together. By using conventional FPM scheme, the construction results lose several information in the phase reconstruction result due to the cells are located at different planes. However, by calculating the defocus distance and implementing our method in subregions 2, the cells missed in the conventional FPM can be recovered effectively as shown in the red arrows.

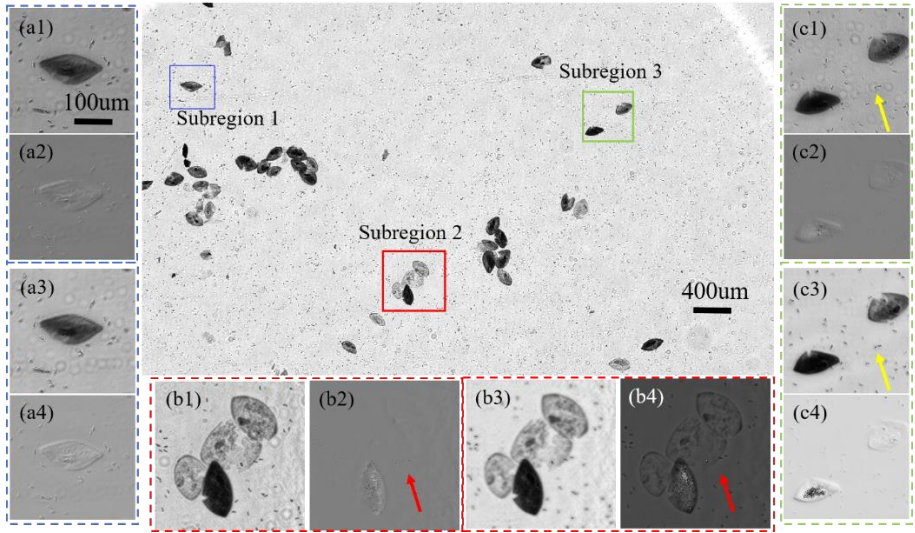

Fig. 9 Phase recovery result of biological sample with the EPRY algorithm and our method. (a1-a4) Reconstructed amplitude and phase of with the EPRY algorithm and our method with 79um defocus distance. (b1-b4) Reconstructed amplitude and phase of with the EPRY algorithm and our method with 80um defocus distance. (c1-c4) Reconstructed amplitude and phase of with the EPRY algorithm and our method with 60um defocus distance.

## 4. Conclusion

Digital refocusing is an efficient strategy of extending DOF in FPM. In this paper, we proposed a fast depth of field extending and digital refocusing FPM. In the optical imaging process, if a sample is placed at out-of-focus plane, there will be an image shift between the center and other BF images corresponding to different illumination angles. This offset is proportional to the defocus distance, in which the coefficient *A* can be calibrated by experiments. Thus, we transform calculating the defocus distance to calculating the image shift between images. Benefited from this strategy, the accurate defocus distances corresponding to different subregions can be obtained mathematically, avoiding the complex optimal process in searching the defocus distances. Then, digital refocusing is utilized in our method to recover the different focused subregions with different defocus distances. Furthermore, by improving the EPRY algorithm with our digital refocusing method, the chromatic aberration coupling with defocus aberration can be recovered correctly. Finally, we use the USAF target and biological samples to demonstrate the feasibility of our method. The DOF of an FPM system can be increased from $\pm 50$ um to more than $\pm 200$ um with a high speed in computation.

Although the actual experimental results have demonstrated the feasibility of our method, some details can be improved in the further works. Firstly, the accuracy of defocus distance can be improved. In our method, defocus distance is obtained by the image shift between the center and other BF images. Although the average of four defocus distances corresponding to two couples of symmetrical BF images can avoid the system misalignments, the accuracy of image registration is still an important factor in calculating the defocus distance. For the further work, more BF images and other image registration methods can be utilized in our method to improve the accuracy of defocus distance. Secondly, some optimal strategies, such as difference map (DM), relaxed averaged alternating reflection (RAAR) and so on, can be used in our method to improve the optimal process.

## Disclosures

The authors declare no conflicts of interest.